\documentclass[aps, prl, amsmath, superscriptaddress,preprintnumbers,twocolumn]{revtex4-2}
\usepackage{graphicx}
\usepackage{dcolumn}
\usepackage{bm}
\usepackage{amsmath}
\usepackage{subfigure}
\usepackage{amsfonts}
\usepackage[svgnames]{xcolor}
\usepackage{longtable}
\usepackage{appendix}
\usepackage{multirow}
\usepackage{makecell}
\usepackage{booktabs}
\usepackage{tabularx}
\usepackage[colorlinks,
            linkcolor=blue,
			filecolor=blue,
			urlcolor= blue,
			citecolor=blue,
            ]{hyperref}
\usepackage{amsthm}
\usepackage{setspace}
\usepackage[T1]{fontenc}
\usepackage{extarrows}

\usepackage[percent]{overpic}
\usepackage{dsfont}
\usepackage{floatrow}
\usepackage{physics}

\newcommand{\ex}[1]{\langle{#1}\rangle}

\newcommand{\blue}[1]{\textcolor{blue}{#1}}
\newcommand{\red}[1]{\textcolor{black}{#1}}

\newcommand{\prlsection}[1]{{\em {#1}.---~}}

\begin{document}
%\begin{CJK*}{UTF8}{gbsn}

\title{Direct characterization of classical dephasing noise for a qubit}
\author{Yuan-De Jin}\thanks{These authors contributed equally to this work}
\affiliation{State Key Laboratory of Semiconductor Physics and Chip Technologies, Institute of Semiconductors, Chinese Academy of Sciences, Beijing 100083, China}
\affiliation{Center of Materials Science and Opto-Electronic Technology, University of Chinese Academy of Sciences, Beijing 100049, China}

\author{Zheng-Fei Ye}\thanks{These authors contributed equally to this work}
\affiliation{State Key Laboratory of Semiconductor Physics and Chip Technologies, Institute of Semiconductors, Chinese Academy of Sciences, Beijing 100083, China}
\affiliation{Center of Materials Science and Opto-Electronic Technology, University of Chinese Academy of Sciences, Beijing 100049, China}
\affiliation{Department of Applied Physics, University of Science and Technology Beijing, Beijing 100083, China}

\author{Wen-Long Ma}
\email{wenlongma@semi.ac.cn}
\affiliation{State Key Laboratory of Semiconductor Physics and Chip Technologies, Institute of Semiconductors, Chinese Academy of Sciences, Beijing 100083, China}
\affiliation{Center of Materials Science and Opto-Electronic Technology, University of Chinese Academy of Sciences, Beijing 100049, China}
\date{\today}
\begin{abstract}

We propose a general method to characterize classical stochastic noise causing qubit dephasing through repetitive Ramsey interferometry measurements (RIMs) on the qubit. Compared to filter-function-based spectroscopy,  our method with simpler pulse control is less constrained by probe coherence time and can directly detect arbitrary-order correlation functions of quasistatic noise processes. We show that each RIM with a short evolution time and suitably chosen control pulses directly samples the noise field and the $n$-point correlations of the RIM outcomes are proportional to the $n$-point correlation functions of the noise processes. We demonstrate the method numerically for two representative cases: an Ornstein-Uhlenbeck Gaussian process and a non-Gaussian ensemble of two-level fluctuators. While practical constraints such as readout contrast and sampling cost persist, our method offers a direct route to quasistatic classical noise spectroscopy across diverse platforms.

\end{abstract}

\maketitle

\prlsection{Introduction} Stochastic processes are widely employed to model fluctuations in various classical or quantum systems \cite{clerk2010a,Milz2021,sinitsyn2016a,Milne2021}. In quantum technologies, the pure dephasing of a qubit as a major form of decoherence is induced by its frequency fluctuations arising from environmental noise \cite{suter2016,yang2017a}. Such frequency fluctuations can often be described as classical noise processes \cite{Ithier2005,szankowski2017,Muller_2019}. For Gaussian noise processes, their statistics is fully captured by the two-point correlation function or noise spectrum \cite{Cywinski2008}. \red{Noise can become intrinsically non-Gaussian in sparse environments containing discrete frequency modes or two-level fluctuators (TLFs) that generate $1/f$ noise \cite{Shnirman2005,galperin2006,galperin2007,Kotler2013,paladino2014}, as well as in nonequilibrium or nonlinearly coupled continuum environments \cite{Uchiyama2002,Maghrebi2016} and more general random telegraph processes \cite{curtis2025}. Higher-order correlations and their polyspectra therefore reveal statistical structure that is inaccessible from the noise spectrum alone. For a TLF ensemble, these quantities can distinguish noise dominated by a few strongly coupled defects from that arising from many comparably contributing fluctuators \cite{Schriefl2006,Wudarski2023,wudarski2023a,Mehmandoost2024}. Incorporating such higher-order statistics into the noise model improves predictions of qubit decoherence under both free and controlled evolution \cite{Mehmandoost2024,norris2016}.
Accurate characterization of these noise processes can therefore guide noise-tailored control strategies \cite{uhrig2007b} and enable nanoscale sensing of environmental dynamics \cite{degen2017a,du2024}.
}
% For sparse environments with discrete frequency modes or two-level fluctuators (TLFs) producing $1/f$ noise \cite{Shnirman2005,galperin2006,galperin2007,Kotler2013,paladino2014}, nonequilibrium or nonlinearly-coupled environments with a continuum of modes \cite{Uchiyama2002,Maghrebi2016}, and random telegraph processes \cite{curtis2025}, the noise statistics can become intrinsically non-Gaussian and require characterization of high-order correlations or noise  polyspectra. 

An extensively used method for classical noise spectroscopy is the filter-function formalism based on dynamical decoupling (DD) control of the qubit \cite{Uys2009,Alvarez2011,yuge2011a,bylander2011,Green2012,PazSilva2014,Soare2014,Ball2015,paz-silva2016,szankowski_accuracy_2018,Biercuk2011,DallaPozza2019,McCourt2023,wang2024,Huang2025,Wang2025}. For Gaussian noise, DD control acts as a tunable filter function to reconstruct the noise spectrum from measured coherence decay traces, which has been demonstrated in diverse platforms, including superconducting qubits \cite{Yan2012,Yan2013,Quintana2017,Gavrielov2025}, semiconductor spin qubits \cite{Cywinski2009a,cywinski2009b,struck2020,kepa2023,Dial2013,Hernandez-Gomez2018,Li2025}, trapped ions \cite{HAFFNER2008,biercuk2009,kotler2011} and molecular qubits \cite{Fu2021}. Extending this framework to non-Gaussian noise requires design of multidimensional frequency comb via sophisticated DD sequences to access noise polyspectra \cite{norris2016,ramon2019,Wang2020,sung2019}. However, the filter-function formalism faces some fundamental limitations. For Gaussian noise spectroscopy, the required long qubit coherence time and pulse sequences make it difficult to characterize low-frequency noise and prone to accumulated pulse errors. For non-Gaussian noise spectroscopy, the complexity of required pulse sequences grows rapidly as the order of detected noise polyspectra increases, demanding even longer qubit coherence time and more precise control \cite{dong2023,dong2025,sung2019}. 

Recent works have proposed to characterize classical noise processes temporally through sequential measurement cycles on a probe qubit \cite{young2012,fink2013a,Sakuldee2019,sakuldee2020b,sakuldee2020a,wudarski2023a,wudarski2023b,Sifft2023}. \red{Because the probe is read out and reset after each short evolution, this approach is less constrained by the qubit coherence time than protocols requiring a single long coherent evolution, and it can achieve arbitrary frequency resolution.} It is therefore particularly suitable for high-resolution magnetic spectroscopy \cite{Laraoui2013,Staudacher2015,Kong2015,boss2017,schmitt2017,glenn2018,yoon2025}, nanoscale nuclear magnetic resonance \cite{Zaiser2016,Pfender2017,Staudenmaier2023,Arunkumar2021,Neuling2023,spohn2025,Maier2025} and nonergodic measurements of quasistatic or low-frequency noise \cite{Sakuldee2019,wudarski2023a}. The two-point and three-point correlators of measurement outcomes for a specific cycle design show some features to distinguish Gaussian and non-Gaussian processes \cite{sakuldee2020a,wudarski2023b}. However, the intrinsic relation between the statistical properties of classical noise processes and the multi-point measurement correlators for generic measurement cycle designs has not been established. \red{Revealing such a relation may help design direct protocols for characterizing quasistatic classical stochastic noise processes.}

\begin{figure}[htbp]
    \centering
    \includegraphics[width=\linewidth]{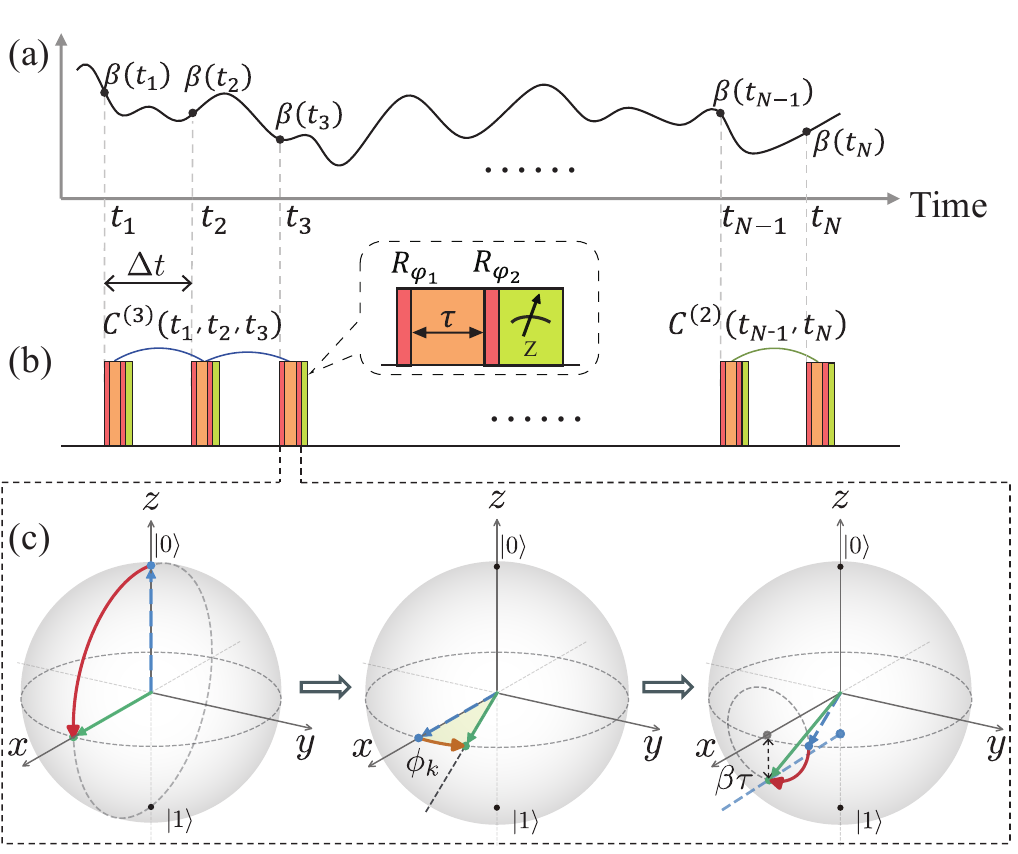}
    \caption{Schematic of a general protocol for characterizing a classical stochastic noise process via sequential RIMs on a qubit. (a) A realization of the continuous stochastic noise process. \red{(b) Direct sampling of the noise field with sequential RIMs. The inset shows one RIM cycle: the qubit with initial state $\ket{0}$, is first rotated by $R_{\varphi_1}$, then evolves freely for a duration $\tau$ while accumulating the phase $\phi_k$, is rotated again by $R_{\varphi_2}$, and is finally measured projectively in the $\sigma_z$-basis. (c) Bloch sphere representation of a RIM before the final measurement. In each sphere, the blue dashed and green solid arrows indicate the qubit state vectors before and after the corresponding step, respectively. The red curves denote two $\pi/2$ rotations, and the orange curve denotes the free precession. The free evolution time $\tau$ in a RIM should be much shorter than the delay time $\Delta t$.  The parameters are $\varphi_1=\pi/2$ and $\varphi_2=\pi$, giving $\Delta\varphi=-\pi/2$. } 
}
    \label{fig:placeholder}
\end{figure}

\red{In this paper, we present a general method to characterize a classical stochastic noise process via a probe qubit. This method is based on sequential Ramsey interferometry measurements (RIMs) on the qubit and can directly detect arbitrary-order quasistatic noise correlations.} The key observation is that the outcome distribution of a RIM with a short evolution time and suitably chosen control pulses constitutes a direct sampling of the noise field, so that the $n$-point correlations of the measurement outcomes are proportional to the $n$-point noise correlations. We numerically illustrate the method for both Gaussian and non-Gaussian stochastic noise processes. Our protocol for measuring the two-point correlation function is efficient in reconstructing the noise spectrum of classical Gaussian noise processes, and its extension to measure higher-order noise cumulants provides a general scheme for non-Gaussian noise characterization.

%but show that higher-order cumulants of the measurement outcomes provide direct signatures of Gaussianity.

%Here we take the $y$-axis as $\varphi_1$ and the $-x$-axis as $\varphi_2$ for the demonstration on the Bloch sphere. 

\prlsection{Basics of classical stochastic noise processes} 
We first introduce the basic concepts of classical stochastic noise processes [Fig. \ref{fig:placeholder}{\color{blue}(a)}]. The pure-dephasing Hamiltonian of a qubit subject to a classical stochastic noise is 
\begin{equation}\label{eq:Ham}
    H=\frac{1}{2}[\Omega+\beta(t)]\sigma_z,
\end{equation}
where $\Omega$ and $\beta(t)$ denote the qubit frequency splitting and fluctuation respectively, and $\sigma_i\,(i=x,y,z)$ is the Pauli-$i$ operator of the qubit. The $n$-point correlation function of $\beta(t)$ is defined as
\begin{equation}
    {C}^{(n)}(t_1,t_2,...,t_n)=\langle\beta(t_1)\beta(t_2)\cdots\beta(t_n)\rangle,
\end{equation}
%or generally the cumulants (irreducible correlations).
where $\langle\cdot\rangle$ denotes the average with respect to the noise realizations, and we assume $\langle \beta(t)\rangle =0$ without loss of generality. In most relevant cases, the stochastic noise processes are time-translation invariant (or stationary), i.e., %$C^{(n)}(t_1,t_2,...,t_n)=C^{(n)}(t_2-t_1,...,t_n-t_1)$. 
$C^{(n)}(t_1,t_2,...,t_n)=C^{(n)}(\Delta\vb*{ t})$ with $\Delta\vb*{ t}=(\Delta t_1,...,\Delta t_{n-1})$ and $\Delta t_{i}=t_{i+1}-t_1$. 
The statistical properties of $\beta(t)$ are fully characterized by the set of irreducible multi-point correlation functions (or cumulants) $\{\tilde{C}^{(n)}(\Delta\vb*{ t})\}$ (see Sec. S1 of Supplemental Materials (SM) for details \footnote{See the Supplemental Material at \url{https://link.aps.org/supplemental/10.1103/6kzz-gld6} for details on correlation functions and cumulants, Monte Carlo estimation of multi-point correlations, reconstruction of correlations with identical time indices, symmetry and higher-order cumulants of TLF noise, effects of nonideal qubit measurements, and experimental acquisition-time estimates for higher-order correlation measurements, which includes Ref.~\cite{Szankowski2023}.}). \nocite{Szankowski2023} We can perform an $(n-1)$-dimensional Fourier transformation for $\tilde C^{(n)}(\Delta\vb*{ t})$ to obtain the $(n-1)$-order noise polyspectrum $S^{(n-1)}(\vb*{\omega})=\int_{\mathbb{R}^{n-1}}e^{-i\vb*{\omega}\cdot\Delta\vb*{ t}}\tilde C^{(n)}(\Delta\vb*{ t})\dd^{n-1}\Delta\vb*{ t}$ with $\vb*\omega=(\omega_1,...,\omega_{n-1})$. Gaussian noise processes have only a second-order cumulant, i.e., all non-zero high-order correlations can be decomposed into two-point correlations. Non-Gaussian noise processes feature at least one additional higher-order cumulant.

\prlsection{Detecting arbitrary noise correlation functions}
We now show that the measurement correlations of repetitive RIMs on the qubit can directly detect arbitrary-order noise correlation functions [Fig. \ref{fig:placeholder}{\color{blue}(b)}] . By transforming to a rotating frame with respect to $H_0=\Omega\sigma_z/2$, we get the effective Hamiltonian $H_I(t)=\beta(t)\sigma_z/2$. We denote $R_\varphi=e^{-i(\cos\varphi\sigma_x+\sin\varphi\sigma_y)\pi/4}$ as the $\pi/2$ rotation with $\varphi$ denoting the rotation
axis in the equatorial plane, and $|\alpha\rangle$ ($\alpha\in\{0,1\}$) as the eigenstate of $\sigma_z$. 

For the $k$th RIM cycle during time $[t_k,t_k+\tau]$, the qubit in initial state $\ket{0}$ undergoes a $\pi/2$ rotation to a superposition state $R_{\varphi_1}\ket{0}=(\ket{0}-ie^{i\varphi_1}\ket{1})/\sqrt{2}$, then a free evolution under $H_I$ for time $\tau$, and finally a second $\pi/2$ rotation $R_{\varphi_2}$ before being projectively measured in the $\sigma_z$-basis [Fig. \ref{fig:placeholder}{\color{blue}(c)}] . 
The probability to obtain outcome $\alpha_k$ is $p(t_k,\alpha_k)=[1+(-1)^{\alpha_k} \cos(\phi_k+\Delta\varphi)]/2$, where $\phi_k=\int_{t_k}^{t_k+\tau}\beta(t)\dd t$ is the accumulated phase during the free evolution and $\Delta\varphi=\varphi_1-\varphi_2$ is the phase difference between two rotation axes. It follows that the expectation value of $\sigma_z$ is
\begin{equation}\label{eq:rk}
    r_k=\langle 0|U_{k}^{\dagger}\sigma_zU_{k}|0\rangle
    =\cos(\phi_k+\Delta\varphi),
\end{equation}
where \red{$U_{k}=R_{\varphi_2}e^{-i\phi_k\sigma_z/2}R_{\varphi_1}$}.
For $n$ RIMs at starting times $\{t_1,t_2,...,t_n\}$, the $n$-point correlation function of the measurement outcomes is
\begin{equation}
    \ex{r_1r_2\cdots r_n}=\Big\langle{\prod_{k=1}^n}\cos(\phi_k+\Delta\varphi)\Big\rangle.
\end{equation}

\red{ We next consider the quasistatic regime, where the noise correlation time $\tau_c$ is much longer than $\tau$, and the noise varies negligibly within each free evolution interval. The accumulated phase can therefore be approximated as $\phi_k=\int_{t_k}^{t_k+\tau}\beta(t)\dd t\approx\beta(t_k)\tau$. For $\beta\tau\ll 1$, we further obtain ${r_k}\approx \cos(\Delta\varphi)[1-\beta^2 (t_k)\tau^2/2]-\sin(\Delta\varphi)\beta (t_k)\tau$. Thus, the phase difference $\Delta\varphi$ serves as a tunable parameter to select different orders of the noise signal.} By choosing $\Delta\varphi=\pm\pi/2$, we have ${r_k}\approx\mp \tau\beta(t_k)$, which directly samples the noise signal at time $t_k$. Consequently, the $n$-point correlations of the RIM outcomes are proportional to the $n$-point correlation functions of the noise processes, so we obtain the main result of our work
\begin{equation}\label{multi}
     \ex{r_1\cdots r_n}\approx\tau^n\ex{\beta(t_1)\cdots\beta(t_n)}=\tau^nC^{(n)}(t_1,...,t_n),
\end{equation}
which holds exactly for $\Delta\varphi=-\pi/2$ and up to an overall factor $(-1)^n$ for $\Delta\varphi=\pi/2$. \red{Thus, Eq.~\eqref{multi} shows that our method directly estimates the multi-time noise correlations $C^{(n)}$. The corresponding cumulants $\tilde C^{(n)}$ are reconstructed algebraically from the estimated correlations using the relations given in Sec.~S1 of the SM~\cite{Note1}, while their $(n-1)$-dimensional Fourier transforms yield the noise polyspectra.
%The relation also holds when we choose $\Delta\varphi=\pi/2$, up to an overall factor $(-1)^n$, i.e., $ \ex{r_1\cdots r_n}\approx (-1)^n\tau^nC^{(n)}(t_1,...,t_n)$, which does not affect the result.
For $n=2$, it provides an efficient method to directly obtain the noise spectrum for Gaussian noise after the Fourier transformation, and for $n=3,4$ it can directly probe the bispectrum and trispectrum for a non-Gaussian noise. }

% We remark that Refs. \cite{young2012,fink2013a,sakuldee2020a,wudarski2023a,wudarski2023b} also proposed to use repetitive RIMs for classical noise characterization, but missed the general relation in Eq. \eqref{multi} because they either only discussed the two-point correlators or adopted $\Delta\varphi=0$ resulting in complicated correlators.
\red{We remark that Refs.~\cite{young2012,fink2013a,sakuldee2020a,wudarski2023a,wudarski2023b} also proposed to use repetitive RIMs for classical noise characterization through finite-time phase accumulation or measurement-induced noise filters. In these approaches, the mapping from measurement correlations to the underlying noise statistics generally depends on the nonlinear phase response or the specific filter functions. Here, we focus on the quasistatic and short-evolution regime, in which this mapping reduces to a direct relation. By choosing $\Delta\varphi=\pm\pi/2$, each RIM serves as a linear sampler of the noise field, so that our protocol can be directly used to construct noise correlations of arbitrary order. Moreover, our method can be regarded as a classical analog of the sequential weak-measurement scheme to characterize arbitrary-order correlations in quantum environments \cite{Liu2010,wang2019a,meinel2022}. The difference here is that classical environments suffer no backaction from qubit measurements, so the detected correlations can be more accurate.}

%Sequential weak measurements have also been utilized to characterize arbitrary-order correlations in quantum environments \cite{wang2019a}.  The difference here is that classical environments suffer no backaction from qubit measurements, so the detected correlators can be more accurate.

%A similar scheme has also been proposed for measuring the noise spectra of generic quantum dephasing environments \cite{Ma2026}. 

%, which is more resource-efficient than DD-based spectroscopy \cite{yuge2011a} and correlation spectroscopy \cite{fink2013a}. 

In practice, we can perform $N$ repetitive RIMs with a delay time $\Delta t$ as a trajectory, and record the measurement outcomes $\{\alpha_k\}_{k=1}^N$. We sample $N_s$ trajectories to obtain the estimated correlation functions (see Sec. S2 of SM \cite{Note1}). 
%$\{\tilde{r}_k\}_{k=1}^N$, where $\tilde{r}_k=\frac{1}{N_s}\sum_{i=1}^{N_s}(-1)^{\alpha_k^{(i)}}$ with $\alpha_k^{(i)}$ being the outcome for the $i$th sample. 
\red{The two-point correlation function can be estimated from the measurement outcome correlations as $C^{(2)}(l_1\Delta t,l_2\Delta t) \approx\frac{1}{\tau^2N_s}\sum_{i=1}^{N_s}(-1)^{\alpha_{l_1}^{(i)}+\alpha_{l_2}^{(i)}}$ with $\alpha_k^{(i)}$ being the outcome in the $i$th trajectory, and $l_k\in\{1,2,...,N\}$ labeling the sampling time $t_{l_k}=l_k\Delta t$.} According to the sampling theorem, the detectable frequency range is $\omega\in[0,\pi/\Delta t]$ gridded by frequency resolution $\pi/(N\Delta t)$. Similarly, the $n$-point correlation can be represented by $C^{(n)}(l_1\Delta t,...,l_{n}\Delta t)\approx \frac{1}{\tau^{n}N_s}\sum_{i=1}^{N_s}\prod_{j=1}^{n}(-1)^{\alpha^{(i)}_{l_j}}$. As the signal amplitude is proportional to $\tau^n$ for the $n$-point correlation, then according to the Hoeffding's inequality, we need $N_s\geq \frac{2}{\delta^2\tau^{2n}}\ln(\frac{2}{\epsilon})$ trajectories to estimate each point in the $n$-point correlation function within an absolute error $\delta$ with probability $1-\epsilon$. \red{Note that the $n$-point correlations for the isolated points containing repeated time indices $l_i=l_j$ cannot be correctly reconstructed by this procedure. However, they can still be obtained by the interpolation methods or constructing a more complicated sequence (see Sec. S3 of SM \cite{Note1})}.

\begin{figure}[htbp]
    \centering
    \includegraphics[width=\linewidth]{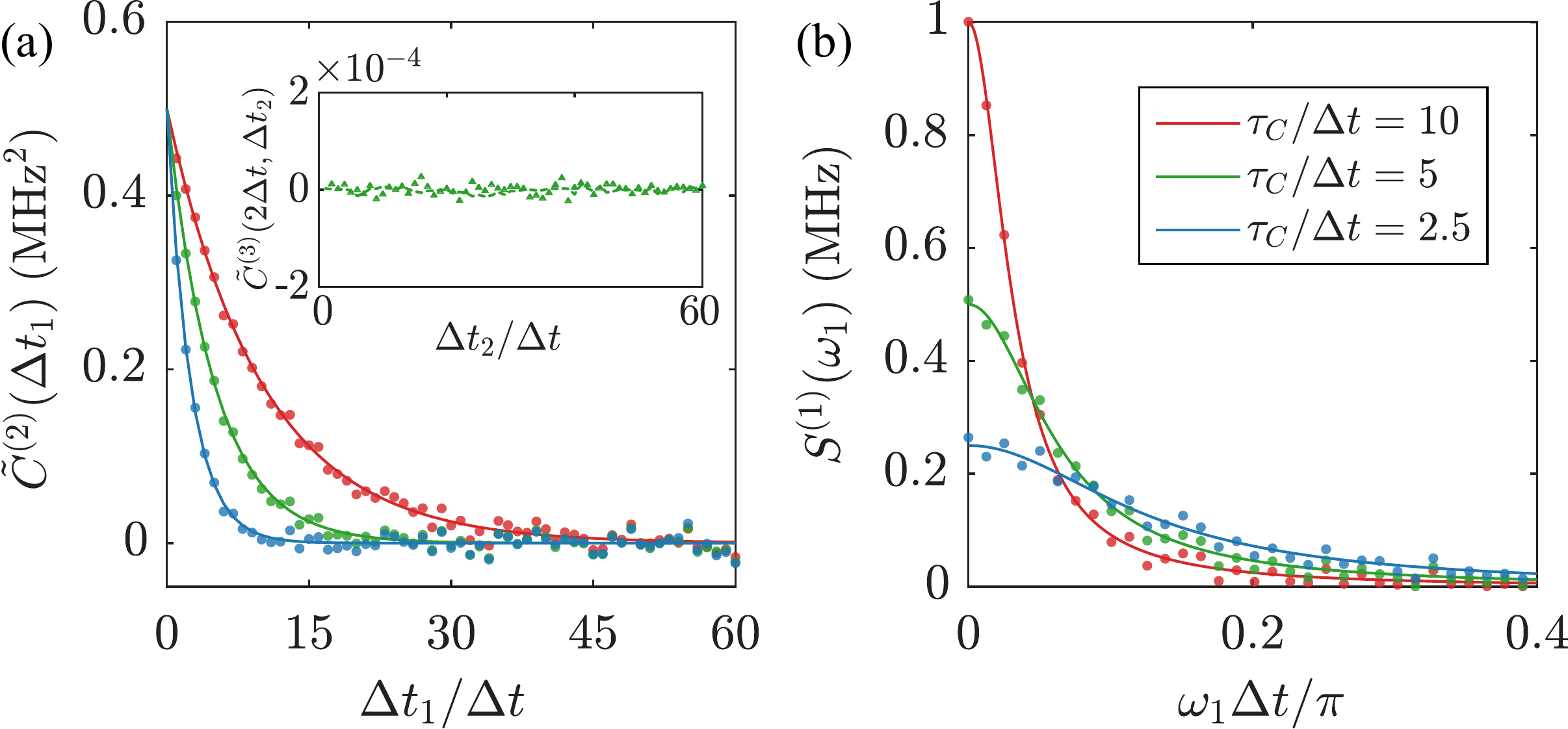}
    \caption{Reconstruction of (a) the second-order cumulant $\tilde{C}^{(2)}(\Delta t_1)$ and (b) noise spectrum $S^{(1)}(\omega)$ of the OU process with three correlation times $\tau_C = 1,\,0.5,\,0.25$ $ \mu$s (red, green, blue line). Solid lines are theoretical results and dots are reconstructed from the RIM outcomes. The inset in (a) shows the third-order cumulant $\tilde{C}^{(3)}(2\Delta t,\Delta t_2)$. The parameters are $\Gamma = 1$ MHz$^2$, $\tau = 0.05$ $\mu$s, $\Delta t = 0.1$ $\mu$s, and $N_s = 5\times 10^8$.}
    \label{fig:ou}
\end{figure}

\prlsection{Example I: Gaussian noise processes}
We first demonstrate our method for the Ornstein-Uhlenbeck (OU) process, the paradigmatic Gaussian noise with nontrivial temporal correlations. The frequency fluctuation $\beta(t)$ in Eq. \eqref{eq:Ham} is governed by the stochastic differential equation 
\red{
\begin{equation}
    \dd \beta=-\gamma\beta \dd t+\sqrt{\gamma\Gamma} \dd W, 
\end{equation}
where $\dd W$ is a Wiener increment satisfying $\langle(\dd W)^2\rangle=\dd t$, $\gamma$ is the spectral width with the correlation time $\tau_C=1/\gamma$, and $\Gamma$ is a noise-strength parameter determining the stationary variance.} In steady state, $\beta(t)$ is a zero-mean Gaussian process with variance $\ex{\beta^2}=\Gamma/2$. Its two-point correlation function and noise spectrum are $\tilde{C}^{(2)}(\Delta t_1)=\Gamma e^{-\gamma|\Delta t_1|}/2$ and $S^{(1)}(\omega_1)=\Gamma\gamma/(\gamma^2+\omega_1^2)$, respectively \cite{uhlenbeck1930}. As a Gaussian process, the OU noise has vanishing cumulants of all orders $n\geq 3$.
We numerically simulate the RIM protocol for OU process with three correlation times. The reconstructed second-order cumulants [Fig.~\ref{fig:ou}\blue{(a)}] and noise spectra [Fig.~\ref{fig:ou}\blue{(b)}] agree well with the exact results for different correlation times, and the third-order cumulant weakly fluctuates around zero due to the sampling errors.

\prlsection{Example II: Non-Gaussian noise processes} As a typical non-Gaussian example, we consider the noise produced by an ensemble of TLFs, which is widely used to model $1/f$ or low-frequency noise \cite{Shnirman2005,paladino2014,wudarski2023a,wudarski2023b}. The $j$th TLF produces a random telegraph noise process by randomly flipping between two states $\{|0\rangle_j,|1\rangle_j\}$ with the switching rate $W_{j}^{+}$ ($W_{j}^{-}$) for the transition $|0\rangle_j\rightarrow|1\rangle_j$ ($|1\rangle_j\rightarrow|0\rangle_j$). The noise field from $N_t$ TLFs is
\begin{equation}
    \beta(t)=\sum_{j=1}^{N_t} \lambda_j[\xi_j(t)-\bar\xi_j],
\end{equation}
where $\lambda_j$ is the coupling strength between the qubit and the $j$th TLF, $\xi_j(t)\in\{1,-1\}$ denotes a classical random telegraph process, and $\bar\xi_j=(W_j^--W_j^+)/W_j$ denotes the asymmetry of the switching rates, with $W_j=W_j^++W_j^-$ being the total switching rate. \red{As the noise from each telegraph process is centered by subtracting $\bar{\xi}_j$, the total noise has zero mean for both symmetric ($\bar{\xi}_j=0$) and asymmetric ($\bar{\xi}_j\neq0$) TLFs. For symmetric TLFs, the stochastic process is invariant under the global sign reversal $\xi_j(t)\to-\xi_j(t)$. Consequently, all odd-order multi-time correlations and cumulants vanish, including the three-time cumulant and the bispectrum. This symmetry alone does not guarantee Gaussianity, as higher even-order cumulants, such as the fourth-order cumulant and trispectrum, can remain nonzero (see Sec.~S4 of SM \cite{Note1}). More generally, an ensemble of independent TLFs can approach a Gaussian process in the large $N_t$ limit, provided that the total noise is not dominated by a small number of exceptionally strongly coupled TLFs. This Gaussian limit does not require the individual TLFs to be symmetric. At finite $N_t$, asymmetry permits nonzero odd-order cumulants, whereas symmetric ensembles may display non-Gaussianity through even-order cumulants. }

%\red{Centering each telegraph process by $\bar{\xi}_j$ makes the total noise zero mean for both symmetric ($\bar{\xi}_j=0$) and asymmetric ($\bar{\xi}_j\neq0$) TLFs. For symmetric TLFs, the stochastic process is invariant under the global sign reversal $\xi_j(t)\to-\xi_j(t)$. Consequently, all odd-order multi-time correlations and cumulants vanish, including the three-time cumulant and the bispectrum. This symmetry alone does not guarantee Gaussianity, higher even-order cumulants, such as the fourth-order cumulant and trispectrum, can remain nonzero. More generally, an ensemble of independent TLFs can approach a Gaussian process in the large $N_t$ limit, provided that the total noise is not dominated by a small number of exceptionally strongly coupled TLFs. This Gaussian limit does not require the individual TLFs to be symmetric. At finite $N_t$, asymmetry permits nonzero odd-order cumulants, whereas symmetric ensembles may retain non-Gaussianity through even-order cumulants.

%More generally, an ensemble of independent TLFs can approach a Gaussian process in the large $N_t$ limit without the requirement of the individual TLFs to be symmetric. At finite $N_t$, asymmetry permits nonzero odd-order cumulants, whereas symmetric ensembles may retain non-Gaussianity through even-order cumulants.
%}

\begin{figure}[htbp]
    \centering
    \includegraphics[width=\linewidth]{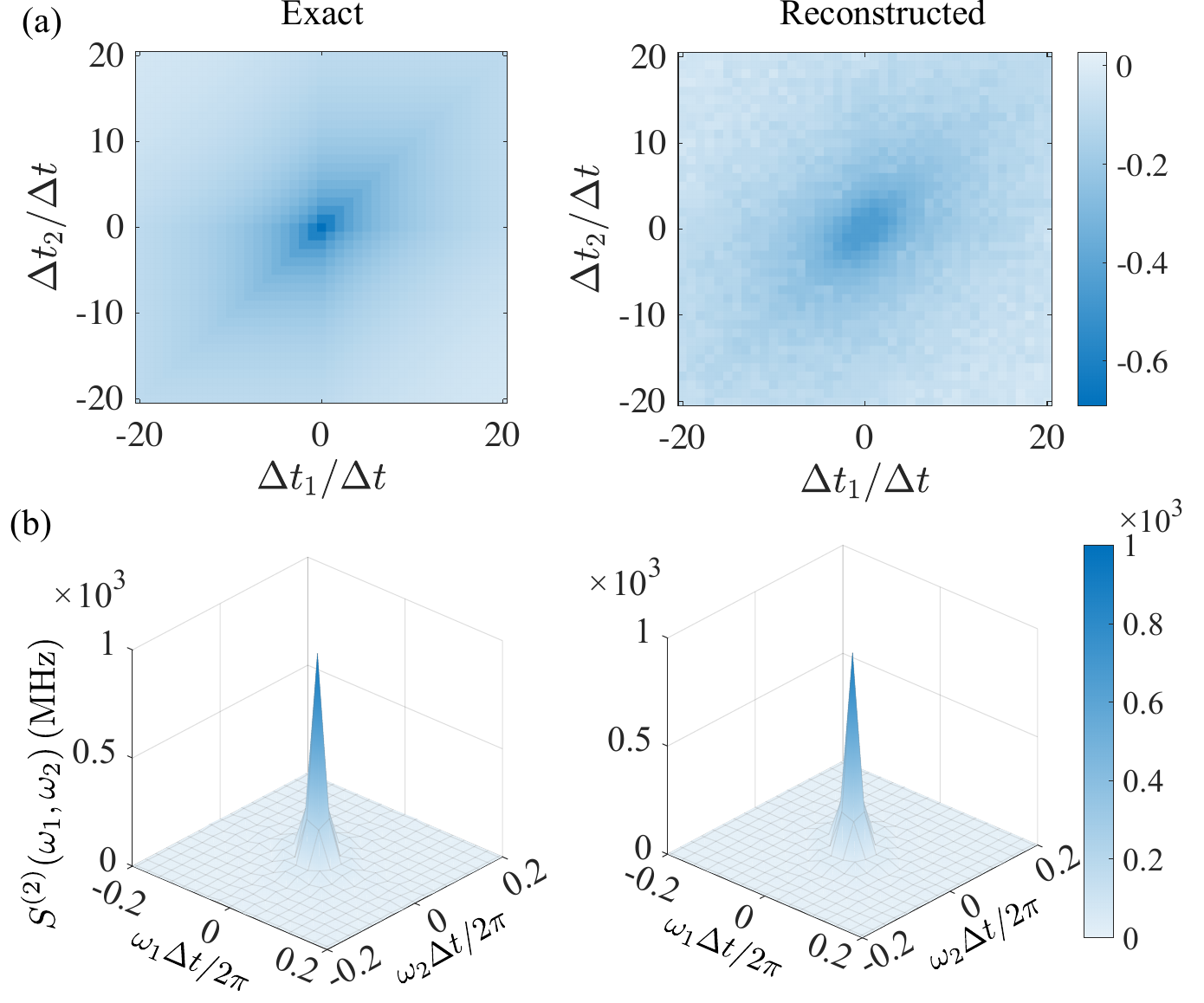}
    \caption{Reconstruction of (a) third-order cumulant $\tilde{C}^{(3)}(\Delta t_1,\Delta t_2)\,(\text{MHz}^3)$ and (b) bispectrum of non-Gaussian noise through our method. The noise is produced by three TLFs with the coupling strength $\lambda=119$ kHz and switching rate $\{W_j\}=\{4.77,21.35,95.49\}$ kHz. Note that the inaccessible points with $\Delta t_1=0$, $\Delta t_2=0$ and $\Delta t_1=\Delta t_2$ for the reconstructed cumulant are smoothed by the neighborhood values. The parameters are $\tau=0.15$ $\mu$s, $\Delta t$=2 $\mu$s and $N_s=4\times 10^8$. }
    \label{fig:tlf3}
\end{figure}
%The exact cumulant and spectrum are normalized by a global factor $\tau^3$.

We first consider a sparse bath containing a few asymmetric TLFs that produces non-Gaussian noise. We show the simulation results to reconstruct the third-order cumulant [Fig.~\ref{fig:tlf3}\blue{(a)}] and the corresponding bispectrum [Fig.~\ref{fig:tlf3}\blue{(b)}] for the noise produced by three TLFs, providing a quantitative characterization of non-Gaussianity. One can see that the cumulant and spectrum from our method agree well with the exact results, and the non-zero third-order cumulants also serve as direct evidence of non-Gaussianity. We also present a simulation of reconstructing fourth-order cumulant and corresponding trispectrum in Sec. S4 of SM \cite{Note1}.

\begin{figure}
    \centering
    \includegraphics[width=0.97\linewidth]{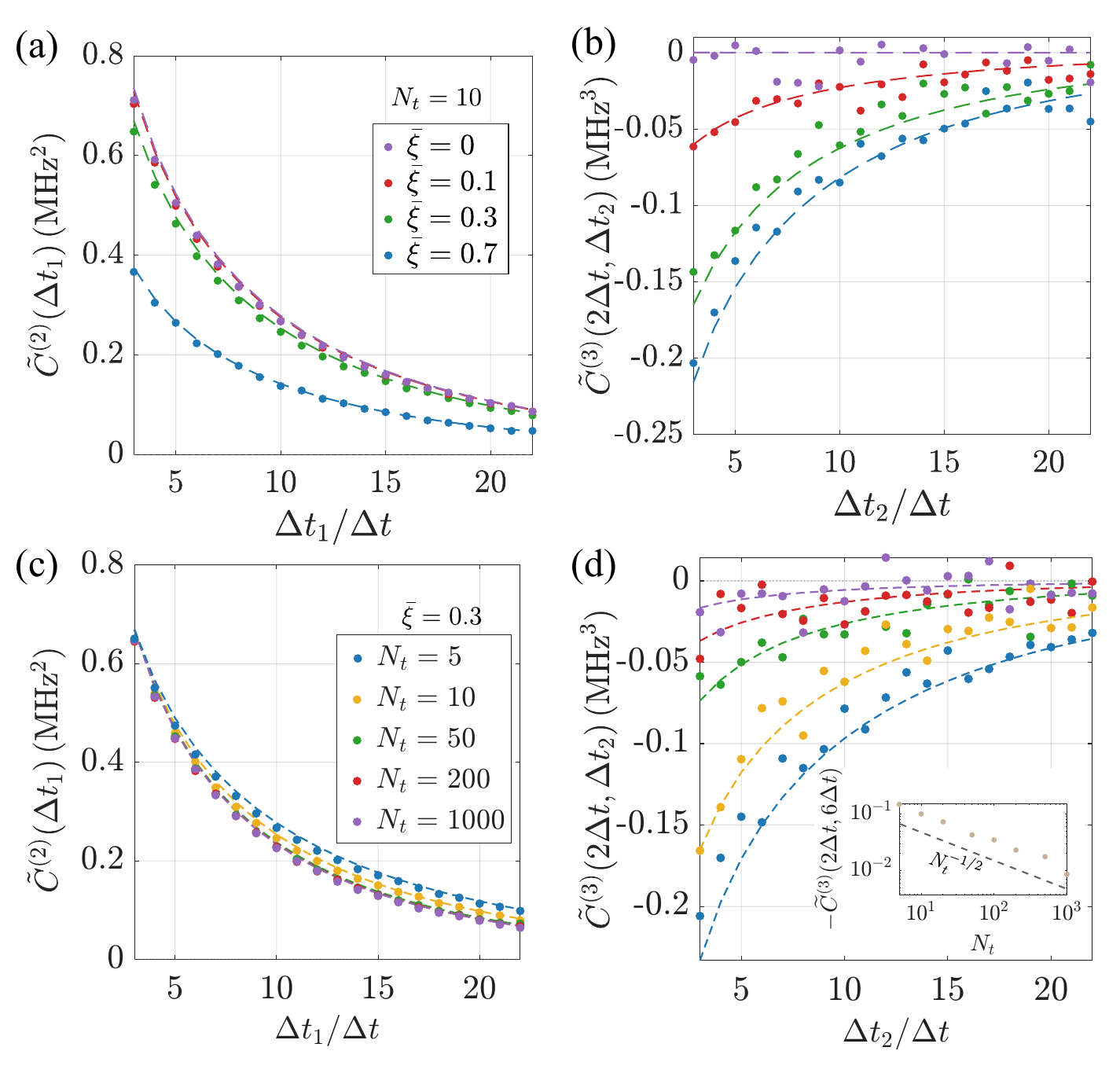}
    \caption{\red{Suppression of the third-order cumulant as the switching-rate asymmetry $\bar\xi$ decreases or the number of TLFs $N_t$ increases. Here we show the ideal (dashed lines) and reconstructed (points) second-order cumulant $\tilde{C}^{(2)}(\Delta t)$ and the slice $\tilde{C}^{(3)}(2\Delta t,\Delta t_2)$ of third-order cumulant with (a-b) different $\bar \xi$ and (c-d) different $N_t$. The coupling strength is $\lambda=207/\sqrt{N_t}$ kHz and the switching rate obeys a log-uniform distribution ranging from 4.77 kHz to 95.49 kHz. The parameters are $\tau=0.15$ $\mu$s, $\Delta t=2$ $\mu$s and $N_s=9\times 10^8$. }}
    \label{fig:10tls}
\end{figure}

\red{We further demonstrate that our method can accurately capture the suppression of third-order noise statistics as either the switching-rate asymmetry decreases or the number of TLFs increases. We first consider $N_t=10$ TLFs with log-uniformly distributed total switching rates $\{W_j\}$ and vary the common switching-rate asymmetry $\bar{\xi}$ from $0.7$ to $0$. As $\bar{\xi}$ decreases, the second-order cumulant increases [Fig.~\ref{fig:10tls}(a)], whereas the third-order cumulant decreases [Fig.~\ref{fig:10tls}(b)]. At $\bar{\xi}=0$, sign-reversal symmetry forces the exact third-order cumulant to vanish, while the reconstructed values fluctuate around zero. We then fix $\bar{\xi}=0.3$ and increase $N_t$ from $5$ to $1000$ with $\lambda_j\propto N_t^{-1/2}$. This normalization keeps the second-order cumulant nearly unchanged [Fig.~\ref{fig:10tls}(c)], whereas the third-order cumulant decays toward zero [Fig.~\ref{fig:10tls}(d)]. The inset confirms the expected $N_t^{-1/2}$ decrease of $\tilde{C}^{(3)}(2\Delta t,6\Delta t)$. Thus, increasing $N_t$ suppresses the non-Gaussianity and shows how an ensemble of independent asymmetric TLFs approaches a Gaussian process in the large-$N_t$ limit. Note that for a clear comparison across different parameter settings, Figs.~\ref{fig:10tls} (b,d) display the same representative nonzero one-dimensional slice, $\tilde C^{(3)}(2\Delta t,\Delta t_2)$, of the full two-dimensional third-order cumulant. }

\prlsection{Experimental considerations} Our method is feasible across a broad range of qubit systems that often require classical noise characterization, such as superconducting qubits, trapped ions, and solid-state spin qubits. %Our method is feasible across a broad class of qubit systems that support repetitive qubit initialization, single-qubit control and readout, such as superconducting qubits, trapped ions, and solid-state spin qubits. 
%Compared to traditional DD-based spectroscopy, our approach offers practical advantages. DD-based methods require long and complex $\pi$-pulse sequences to access higher-order polyspectra, which makes them susceptible to accumulated control errors and strictly limits their application to noise frequencies bounded by the qubit lifetime. In contrast, each cycle in our method only requires a short Ramsey interferometry and readout, which is more feasible for practical systems. 
Compared with conventional DD-based spectroscopy, our method replaces long and increasingly complicated $\pi$-pulse sequences by short Ramsey cycles followed by repetitive readout and reset, resulting in highly tunable bandwidth and spectral resolution. \red{The operating conditions are the quasistatic regime $\tau\ll\tau_c$ and the short-evolution condition $\beta\tau\ll1$.} Thus, our method is especially attractive for detecting low-frequency noise and higher-order polyspectra, where pulse errors and sequence complexity become increasingly challenging for traditional DD-based methods. 

In realistic experiments, qubit decoherence, measurement errors, and finite pulse and readout times can reduce the efficiency of our method by reducing the signal amplitude. 
For example, a $T_2$-type noise reduces the readout contrast, adding a damping factor to Eq.~\eqref{eq:rk} as $r_k=e^{-\tau/T_2}\cos(\phi_k+\Delta\varphi)$. Since it appears as an overall factor, and the evolution time $\tau$ is short, it does not affect our main result but only increases the required sample number. A finite readout and reset time $t_{d}$ can also be incorporated straightforwardly by replacing the ideal sampling interval $\Delta t$ with the actual cycle time $\Delta t+t_d$, which just alters the frequency window and resolution. \red{Accordingly, the upper bound of the detectable frequency becomes $\omega_{\max{}}=\pi/(\Delta t+t_d)$ and approaches the theoretical upper limit $\pi/(\tau+t_d)$ for sequential RIMs without a wait time, subject to the quasistatic condition $\omega\tau\ll1$.}

%, and would not be accumulated in the pulse sequences. 

%Furthermore, our scheme is highly compatible with platforms where the readout is weak or non-projective, 
Furthermore, our scheme is also applicable to nonideal qubit measurements, such as weak optical readout for solid-state defect spins or trapped ions and imperfect projective measurements for superconducting qubits (see Sec.~S5 of the SM \cite{Note1} for details). In the former case, a single readout may yield much less than one detected photon on average, preventing a single-shot projective measurement. Such a readout is more naturally described by the Kraus operators
$K_n=\sum_{\alpha=0,1}\sqrt{p(n|\alpha)}P_\alpha$,
where $p(n|\alpha)$ is the probability of collecting $n$ photons in one readout when the probe is in the state
$P_\alpha=\ket{\alpha}\bra{\alpha}$.
When the average photon number is small, the readout outcomes can be truncated to $K_0$ and $K_1$, with
$p(1|\alpha)=g_\alpha$ and $p(0|\alpha)=1-g_\alpha$.
The probability of detecting a photon in the $k$th RIM is therefore
\begin{math}
    \ex{g_k}=\bar g+\Delta r_k,
\end{math}
where $\bar g=(g_0+g_1)/2$ and $\Delta=(g_0-g_1)/2$.
Thus, multi-point photon-count correlations can be used to estimate the corresponding multi-point noise correlations
\begin{math}
    \ex{\delta g_1\cdots\delta g_n}=\tau^n\Delta^n C^{(n)}(t_1,...,t_n),
\end{math}
 with $\delta g_k=g_k-\bar g$.
Thus, even when the optical readout is not fully projective, the photon-count record can still serve as a practical estimator of the classical noise correlations. Imperfect projective measurements lead to a similar reduction of the signal contrast.

\red{These reductions in signal contrast also limit the detected correlation order in practice. We define the dimensionless measurement signal amplitude $A_n=|\tau^n C^{(n)}(t_1,\ldots,t_n)|$ and an effective contrast factor $G\leq1$, such that the measured $n$-point signal amplitude is reduced to $G^n A_n$. For a relative estimation error $\eta$, the required sample number can be written as
\begin{equation}
    N_s \geq \frac{2}{\eta^2G^{2n}A_n^2}\ln\left(\frac{2}{\epsilon}\right).
\end{equation}
For unit contrast $G=1$, taking $\beta\tau\sim0.1$-$0.25$, $\eta=0.1$, and $1-\epsilon=0.95$ gives approximately $N_s\sim2\times10^5$-$7\times10^6$, $3\times10^6$-$7\times10^8$, and $5\times10^7$-$7\times10^{10}$ for $n=2$, $3$, and $4$, respectively. A finite contrast increases these values by a factor $G^{-2n}$, making the effect of readout imperfections progressively more important at higher orders. We estimate the total acquisition time using representative control and readout parameters for superconducting qubits \cite{Xiong2026,Li2023} and NV centers with high-contrast readout at low temperature \cite{Abobeih2018,Bartling2025,Blok2014,Bonato2016} (see Sec.~S6 of the SM \cite{Note1}). These estimates indicate that the detection of up to the fourth-order correlations is experimentally feasible in both platforms, although the four-point correlation detection requires considerably longer acquisition time. 
}

\prlsection{Conclusions and outlooks} We have presented a general protocol to characterize classical stochastic noise processes via repetitive RIMs on a single qubit. We find that in the short-evolution regime and with suitable pulse control, each RIM can directly sample the noise field, and thus $n$-point correlation functions of the measurement outcomes are proportional to the $n$-point noise correlations. This establishes a theoretical framework to efficiently reconstruct the noise spectrum and polyspectrum from sequential measurements, while avoiding the rapidly increasing pulse-sequence complexity of conventional DD-based spectroscopy. 

The present framework can be extended in several directions. \red{First, the present protocol assumes uniformly spaced temporal sampling, while it is desirable to develop more efficient data acquisition strategies to reduce the overall sampling overhead \cite{Ferrie2018,Dinani2019}, such as nonuniform temporal sampling, quantum Bayesian methods, and measurement-based adaptive feedback.} Second, it can be extended to characterize a quantum environment after incorporating the measurement backaction \cite{kuffer2022,kuffer2025,machado2023,wang2021a}. Additionally, extensions to multiple probes may further enable spatiotemporal characterization of non-local correlated classical or quantum noise processes \cite{vonlupke2020,Hainzer2024,Xia2025,brady2026}.%Our method is feasible for a broad class of present-day experimental systems.

%\red{First, the sample-complexity estimates presented above assume the uniformly spaced temporal sampling adopted in the present protocol, more resource-efficient acquisition strategies, such as nonuniform temporal sampling, quantum Bayesian methods, and measurement-based adaptive feedback, may further reduce the overall sampling overhead \cite{Ferrie2018,Dinani2019}, developing and benchmarking such optimized protocols constitutes an interesting direction for future work.} 

%First, we expect that Bayesian algorithm and adaptive control may further make our method more resource-efficient by extracting maximal information from each qubit measurement \cite{Ferrie2018,Dinani2019}.

%\newpage
\prlsection{Acknowledgments} We thank Yi-Xu Wang for valuable discussions. The research is supported by the National Natural Science Foundation of China (No. 12174379, No. 12574082, No. E31Q02BG), the Chinese Academy of Sciences (No. E0SEBB11, No. E27RBB11), Quantum Science and Technology-National Science and Technology Major Project (No. 2021ZD0302300) and Chinese Academy of Sciences Project for Young Scientists in Basic Research (YSBR-090).

\textit{Data availability.} The data that support the findings of this article are openly available \footnote{Y.-D. Jin, Z.-F. Ye and W.-L. Ma, Data for paper ``Direct characterization of classical dephasing noise for a qubit'', Zenodo, 2026, \url{https://doi.org/10.5281/zenodo.19879458}}.

 \bibliography{ref}
% \clearpage
% \begin{figure}
%     \centering
%     \includegraphics[width=\linewidth]{1.pdf}
%     \caption{Caption}
%     \label{fig:placeholder}
% \end{figure}

\end{document}

%\begin{CJK*}{UTF8}{gbsn}
%\linenumbers

%\title{Characterizing a Classical Stochastic Process via a Single Qubit}

%\title{Universal Characterization of Classical Stochastic Noise Processes}
\title{Supplemental Materials for ``Direct characterization of classical dephasing noise for a qubit''}
\author{Yuan-De Jin}\thanks{These authors contributed equally to this work}
\affiliation{State Key Laboratory of Semiconductor Physics and Chip Technologies, Institute of Semiconductors, Chinese Academy of Sciences, Beijing 100083, China}
\affiliation{Center of Materials Science and Opto-Electronic Technology, University of Chinese Academy of Sciences, Beijing 100049, China}

\author{Zheng-Fei Ye}\thanks{These authors contributed equally to this work}
\affiliation{State Key Laboratory of Semiconductor Physics and Chip Technologies, Institute of Semiconductors, Chinese Academy of Sciences, Beijing 100083, China}
\affiliation{Center of Materials Science and Opto-Electronic Technology, University of Chinese Academy of Sciences, Beijing 100049, China}
\affiliation{Department of Applied Physics, University of Science and Technology Beijing, Beijing 100083, China}

\author{Wen-Long Ma}
\email{wenlongma@semi.ac.cn}
\affiliation{State Key Laboratory of Semiconductor Physics and Chip Technologies, Institute of Semiconductors, Chinese Academy of Sciences, Beijing 100083, China}
\affiliation{Center of Materials Science and Opto-Electronic Technology, University of Chinese Academy of Sciences, Beijing 100049, China}
\date{\today}
\begin{abstract}

\end{abstract}
\maketitle

\tableofcontents
\section{Basics of correlations and cumulants}
We first introduce some preliminaries about the relations between the correlation functions and cumulants of classical qubit dephasing noise \cite{norris2016,ramon2019,Wang2020,sung2019}. In the main text, we have defined the $n$-point correlation function as $n$th-order moment of the stochastic noise
\begin{equation}
    C^{(n)}(t_1,\ldots,t_n)=\ex{\beta(t_1)\cdots\beta(t_n)}.
\end{equation}
The cumulant is defined through the cumulant generating function
\begin{equation}
    \tilde C^{(n)}(t_1,\ldots,t_n)=\left.\frac{\partial^n}{\partial k_1\cdots\partial k_n}\ln \left\langle{\exp\left[\vb*{k\cdot\beta}\right]}\right\rangle\right|_{ \vb*{k}=0},
\end{equation}
where $\vb*{k}=(k_1,...,k_n)$ is an auxiliary variable, $\vb*\beta=(\beta(t_1),...,\beta(t_n))$. The correlations and cumulants are related by the partition formula
\begin{equation}
    C^{(n)}(t_1,\ldots,t_n)=\sum_{\mathcal{P}\in\Pi_n}\prod_{B\in\mathcal{P}}\tilde C^{(|B|)}(\{t_i\}_{i\in B}),
\end{equation}
where $\Pi_n$ is the set of all partitions of $\{1,\ldots,n\}$, $\mathcal{P}$ denotes one partition, $B$ is a block in that partition, and $|B|$ is the number of elements in the block. This relation shows that the correlation functions can be constructed by all possible cumulants. For example, for zero-mean noise, $\tilde C^{(2)}=C^{(2)}$ and $\tilde C^{(3)}=C^{(3)}$, and
\begin{equation}
    \tilde C^{(4)}(t_1,t_2,t_3,t_4)=C^{(4)}(t_1,t_2,t_3,t_4)
    -C^{(2)}(t_1,t_2)C^{(2)}(t_3,t_4)
    -C^{(2)}(t_1,t_3)C^{(2)}(t_2,t_4)
    -C^{(2)}(t_1,t_4)C^{(2)}(t_2,t_3).
\label{C4}
\end{equation}

\

\section{Details in Monte Carlo sampling}
Here we illustrate how the multi-point measurement correlations can be estimated by the Monte Carlo sampling over the classical noise realizations \cite{sakuldee2020a,wudarski2023b}. The projective measurement in the $Z$-basis corresponds to an observable 
\begin{equation}
    \sigma_z=\sum_{\alpha=0,1}(-1)^\alpha P_\alpha,
\end{equation}
where $P_\alpha=\ket{\alpha}\bra{\alpha}$ is the projector to the subspace. \red{Since the qubit is periodically reset and the environment is classical, the probability of obtaining a sequence of outcomes $\{\alpha_1,..., \alpha_n\}$ after $n$ RIMs at time $\{t_1,...,t_n\}$ in a fixed classical noise realization is
\begin{equation}
    p(t_1,\alpha_1;...;t_n,\alpha_n)=\prod_{k=1}^n p(t_k,\alpha_k),
\end{equation}
with $p(t_k,\alpha_k)$ being the probability of obtaining outcome $\alpha_k$ for the $k$th RIM. }We can derive $p(t_k,\alpha_k)$ directly as
\begin{equation}
    p(t_k,\alpha_k)=\Tr[P_\alpha U_k\ket{0}\bra{0} U_k^\dagger ]=[1+(-1)^{\alpha_k}\sin(\phi_k)]/2\approx [1+(-1)^{\alpha_k}\beta(t_k)\tau]/2,
\end{equation}
where we have set $\Delta\varphi=-\pi/2$ and assumed $\tau\ll 1$. So the measurement correlation function can be expressed as
\begin{equation}\label{eq:r1rn}
\begin{aligned}
    \ex{r_1\cdots r_n}&=\ex{\sum_{\{\alpha_1,...,\alpha_n\}}p(\alpha_1,...,\alpha_n)\prod_{k=1}^{n}(-1)^{\alpha_k}}\\
        &=\ex{\prod_{k=1}^n\sum_{\alpha_k=0,1}p(t_k,\alpha_k)(-1)^{\alpha_k}}=\tau^n\ex{\beta(t_1)\cdots \beta(t_n)},
\end{aligned}
\end{equation}
where $\ex{}$ denotes the average over the noise realizations.
We sample a sufficient number of trajectories $N_s$, and record the specific measurement outcome sequence $\{\alpha_1^{(i)},...,\alpha_n^{(i)}\}_{i=1}^{N_s}$.  Therefore, the $n$-point measurement correlation function can be estimated from the trajectories
\begin{equation}
     \ex{r_1\cdots r_n}\approx\frac{1}{N_s}\sum_{i=1}^{N_s}\prod_{k=1}^n(-1)^{\alpha_k^{(i)}},
\end{equation}
and thus
\begin{equation}
    C^{(n)}(t_1,...,t_n)\approx\frac{1}{\tau^n N_s}\sum_{i=1}^{N_s}\prod_{k=1}^n(-1)^{\alpha_k^{(i)}}.
\end{equation}
Note that a complete estimation for Eq.~\eqref{eq:r1rn} requires two layers of averaging: averaging over the projective measurement outcomes conditioned on a fixed classical noise realization, and the ensemble averaging over different classical noise realizations. In practice, we can seldom obtain multiple sets of measurement results on a certain noise realization. However, for a sufficiently large number of independent trajectories, the Monte Carlo samples still provide a reliable estimate of the desired measurement correlation as shown in the main text. 

% \begin{equation}
% \begin{aligned}
%         \hat C^{(n)}(t_1,...,t_n)&=\frac{1}{N_s}\sum_{i=1}^{N_s}p_i(\alpha_1,...,\alpha_n)\prod_{k=1}^{n}(-1)^{\alpha_k^{(i)}}\\
%         &=\frac{1}{N_s}\sum_{i=1}^{N_s}\prod_{k=1}^n (-1)^{\alpha_k^{(i)}}p_i(t_k,\alpha_k)
% \end{aligned}
% \end{equation}

\section{Estimation of correlation functions with identical time indices}
The method proposed in the main text is designed for estimating correlation functions with different time indices, but is invalid when the correlation functions have identical time indices. For example, if two indices coincide in a three-point correlation function, i.e., $C^{(3)}(t_j,t_k,t_p)$ with $t_j=t_k$, the measurement outcome $\alpha_{k}^{(i)}$ in the $i$th trajectory (or noise realization) obeys $(-1)^{\alpha_{j}^{(i)}+\alpha_{k}^{(i)}}=(-1)^{2\alpha_{k}^{(i)}}=1$. So the measurement product no longer samples the corresponding multi-point correlation $\ex{\beta^2(t_j)\beta(t_p)}$ but just $\ex{\beta(t_p)}$, which produces zero for a zero-mean noise. Therefore, the points with repeated indices, such as $t_j=t_k$, $t_k=t_p$, or $t_j=t_p$ in a three-point correlation function, cannot be correctly estimated by this reconstruction procedure. 

%While, the limitation does not affect the identification of Gaussianity nor the overall reconstruction effect, as shown in the main text. Although it is not significant for the main result, 

There are two ways to solve this problem. First we note that these points are isolated and therefore can be easily obtained by the interpolation methods, as shown in Fig. 3 of the main text. Alternatively
these points can still be estimated through adjusting the measurement protocol. According to the main text, the expectation value of $\sigma_z$ in $k$th single RIM is 
\begin{equation}
    r_k=\cos(\phi_k+\Delta\varphi)\approx \cos(\Delta\varphi)[1-\beta^2 (t_k)\tau^2/2]-\sin(\Delta\varphi)\beta (t_k)\tau,
\end{equation}
where $\phi_k$ is the accumulated phase in the evolution and $\Delta\varphi$ is the phase difference between two $\pi/2$-pulses. When we choose $\Delta\varphi=-\pi/2$, we obtain the linear response of the noise with
\begin{equation}
    r_k^{l}\approx\tau \beta(t_k),
\end{equation}
which is used in the main text. While we can also tune $\Delta\varphi=\pi$ to obtain the quadratic response with
\begin{equation}
    r_k^{q}\approx\tau^2\beta^2 (t_k)/2-1.
\end{equation}
When choosing $\Delta \varphi=\pi$ for the $j$th RIM and $\Delta \varphi=-\pi/2$ for the $p$th RIM, we have
\begin{equation}
    \ex{(r_j^q+1)r_p^l}\approx\tau^3\ex{\beta^2(t_j)\beta(t_p)}/2.
\end{equation}
So generally, whenever a correlation function contains repeated time indices, one can replace the two RIMs with $\Delta\varphi=-\pi/2$ at the same time by a single RIM with $\Delta \varphi=\pi$.

\begin{figure}
    \centering
    \includegraphics[width=0.8\linewidth]{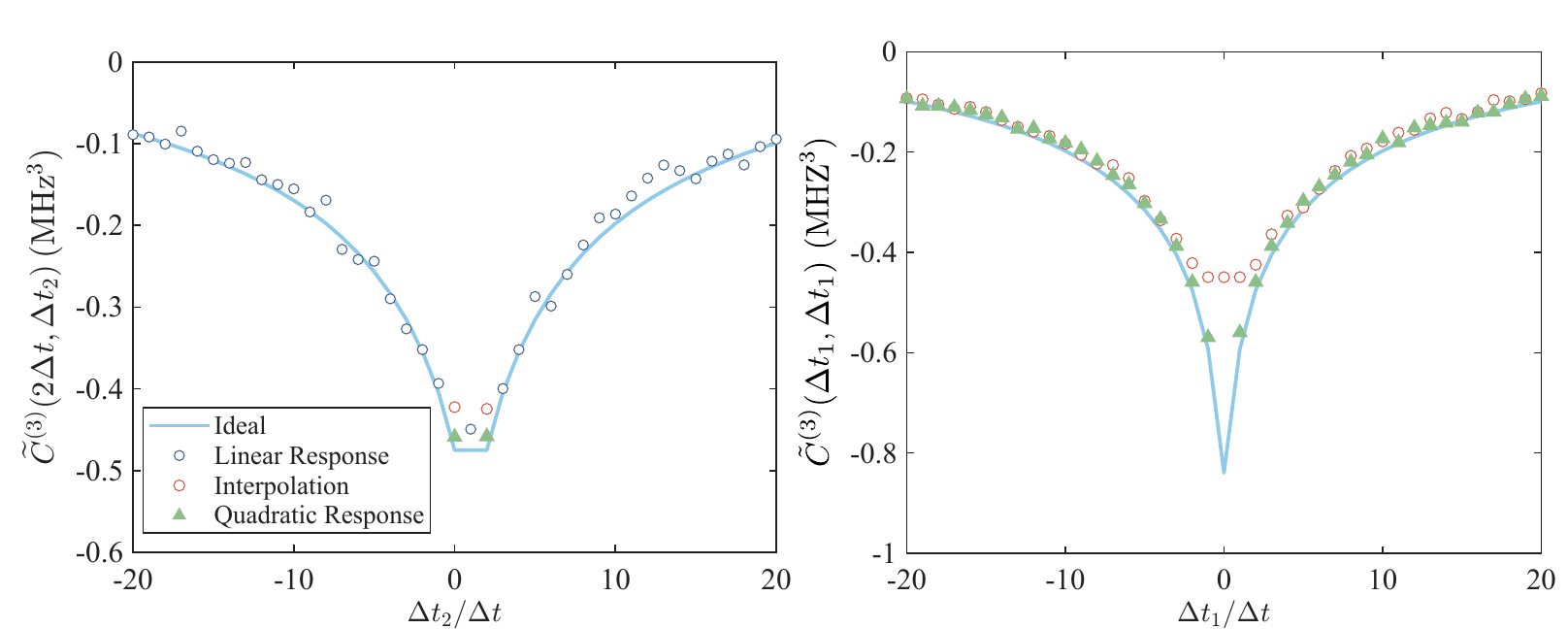}
    \caption{\red{Reconstruction of points with identical time indices in two slices of the third-order cumulant $\tilde C^{(3)}(2\Delta t,\Delta t_2)$ and $\tilde C^{(3)}(\Delta t_1,\Delta t_1)$. The quadratic response protocol (green triangles) accurately follows the exact result (solid line), while interpolation (red circles) reproduces the overall trend over a broad range of time delays. The model and parameters are the same as those used in Fig.~3 in the main text.}}
    \label{fig:identical_indices}
\end{figure}

\red{Figure~\ref{fig:identical_indices} provides a numerical illustration of this alternate protocol and compares it with the interpolation used in Fig.~3 of the main text. We consider two representative slices of the third-order cumulant. For $\tilde C^{(3)}(2\Delta t,\Delta t_2)$, repeated indices occur at $\Delta t_2=0$ and $\Delta t_2=2\Delta t$, whereas the diagonal slice $\tilde C^{(3)}(\Delta t_1,\Delta t_1)$ fully consists of repeated indices. The quadratic response protocol can accurately follow the exact cumulant for these points. Interpolation also captures the overall dependence over a broad range of time delays, although it smooths the sharp feature near the center of the diagonal slice.}

\red{
%\section{Correlation functions of two-level fluctuators}
\section{Symmetry and cumulants of TLF noise}
In this section, we analyze the symmetry properties and higher-order cumulants of two-level fluctuator (TLF) noise. We first consider the $j$th TLF introduced in Eq.~(7) of the main text. The telegraph variable $\xi_j(t)\in\{+1,-1\}$ takes the value $+1$($-1$) when the TLF occupies the state $|0\rangle_j$($|1\rangle_j$). We further define the centered telegraph variable $\delta\xi_j(t):=\xi_j(t)-\bar{\xi}_j$, and we have $\langle\delta\xi_j(t)\rangle=0$. The transition rates $|0\rangle_j\to |1\rangle_j$($|1\rangle_j\to |0\rangle_j$) are $W_j^+$($W_j^-$), with the total switching rate $W_j=W_j^++W_j^-$. 
%We denote the stationary occupation probabilities of the two states by $p_{i,j}$, $i\in\{0,1\}$. The balance equation gives $p_{i,j}=W_j^\pm/W_j$ for $i=1,0$, respectively. The stationary mean of the telegraph variable is therefore
%\begin{equation}
%\langle\xi_j(t)\rangle=p_{0,j}-p_{1,j}=\frac{W_j^--W_j^+}{W_j}\equiv \bar{\xi}_j,
%\end{equation}
%which is the asymmetry parameter defined in Eq.~(7) of the main text. We further define the centered telegraph variable $\delta\xi_j(t)=\xi_j(t)-\bar{\xi}_j$, and we have $\langle\delta\xi_j(t)\rangle=0$. The contribution of this TLF to the total noise field is $\beta_j(t)=\lambda_j\delta\xi_j(t)$, Eq.~(7) can therefore be written as $\beta(t)=\sum_{j=1}^{N_t}\beta_j(t)$. Note that the centering makes the noise zero mean for both symmetric and asymmetric TLFs.
}

\red{
The telegraph process is Markovian. For a stationary TLF, its conditional statistics at a later time depend only on its present state \cite{Szankowski2023}. In particular, for a stationary TLF and $\Delta t\geq0$, it satisfies
\begin{equation}
\mathbb{E}\!\left[\xi_j(t+\Delta t)\,\middle|\,\xi_j(t)\right]=\bar\xi_j+
\left[\xi_j(t)-\bar\xi_j\right]e^{-W_j\Delta t}
\end{equation}
where $\mathbb{E}[\cdot\,|\,\cdot]$ denotes the conditional expectation. Equivalently,
\begin{equation}
\mathbb{E}\!\left[\delta\xi_j(t+\Delta t)\,\middle|\,\delta\xi_j(t)\right]=e^{-W_j\Delta t}\delta\xi_j(t).
\label{eq:conditional-expectation}
\end{equation}
The stationary variance is
\begin{equation}
\left\langle\delta\xi_j^2(t)\right\rangle=\left\langle\left[\xi_j(t)-\bar\xi_j\right]^2\right\rangle=1-\bar\xi_j^2.
\label{eq:propagatorcentered}
\end{equation}
Using Eq.~\eqref{eq:conditional-expectation}, the two-time correlation function is
\begin{equation}
\left\langle\delta\xi_j(t_1)\delta\xi_j(t_2)\right\rangle=\left(1-\bar\xi_j^2\right)e^{-W_j|t_2-t_1|}.
\label{C2}
\end{equation}
}

\red{
We denote the contribution of the $j$th TLF to the total noise field by
$\beta_j(t):=\lambda_j\delta\xi_j(t)$, so that Eq.~(7) of the main text becomes
$\beta(t)=\sum_{j=1}^{N_t}\beta_j(t)$. Since $\langle\beta_j(t)\rangle=0$, its
second-order correlation function is identical to its second-order cumulant
\begin{equation}
\tilde C^{(2)}_{\beta_j}(t_1,t_2)=\lambda_j^2\left(1-\bar\xi_j^2\right)e^{-W_j|t_2-t_1|},
\end{equation}
here we use $\tilde C_{\beta_j}^{(n)}$ to denote the $n$th-order cumulant of $\beta_j$. The dependence of the higher-order statistics on the switching rate asymmetry becomes
explicit at third order. We next derive a recursion relation for higher-order correlation functions. Since $\xi_j^2(t)=1$, the centered variable satisfies
\begin{equation}
\delta\xi_j^2(t)=1-\bar\xi_j^2-2\bar\xi_j\delta\xi_j(t).
\label{eq:deltaxi}
\end{equation}
For ordered times $t_1\leq\cdots\leq t_n$, Eq.~\eqref{eq:conditional-expectation} and Eq.~\eqref{eq:deltaxi} give 
\begin{equation}
\left\langle\delta\xi_j(t_1)\cdots\delta\xi_j(t_n)\right\rangle=
-2\bar\xi_j e^{-W_j\Delta t_{n-1,n}}\left\langle\delta\xi_j(t_1)\cdots\delta\xi_j(t_{n-1})\right\rangle+\left(1-\bar\xi_j^2\right)e^{-W_j\Delta t_{n-1,n}}
\left\langle\delta\xi_j(t_1)\cdots\delta\xi_j(t_{n-2})\right\rangle,
\label{eq:recursion}
\end{equation}
where $\Delta t_{n-1,n}:= t_n-t_{n-1}$. As the simplest nontrivial application, setting $n=3$ in Eq.~\eqref{eq:recursion} gives
\begin{equation}
\left\langle\delta\xi_j(t_1)\delta\xi_j(t_2)\delta\xi_j(t_3)\right\rangle=
-2\bar\xi_j\left(1-\bar\xi_j^2\right)e^{-W_j(t_3-t_1)},\qquad t_1\leq t_2\leq t_3.
\end{equation}
Thus, 
\begin{equation}
\tilde C_{\beta_j}^{(3)}(t_1,t_2,t_3)=-2\lambda_j^3\bar\xi_j\left(1-\bar\xi_j^2\right)e^{-W_j(t_3-t_1)},\qquad t_1\leq t_2\leq t_3.
\end{equation}
The third-order cumulant is therefore generally nonzero for an asymmetric TLF and vanishes for a symmetric TLF.  Moreover, it depends only on the outer time span $t_3-t_1$ rather than the intermediate time $t_2$.
}
\red{
For a symmetric TLF, Eq.~\eqref{eq:recursion} reduces to
\begin{equation}
\left\langle\delta\xi_j(t_1)\cdots\delta\xi_j(t_n)\right\rangle_{\bar\xi_j=0}=e^{-W_j\Delta t_{n-1,n}}\left\langle\delta\xi_j(t_1)\cdots\delta\xi_j(t_{n-2})\right\rangle_{\bar\xi_j=0}.
\label{eq:recursion-symmetric}
\end{equation}
The recursion reduces the order by two at each step. For odd $n$, it terminates at $\langle\delta\xi_j(t_1)\rangle=0$, showing that all odd-order correlation functions vanish exactly. The moment-cumulant relation implies recursively that all odd-order cumulants vanish as well, hence
\begin{equation}
\tilde C_{\beta_j}^{(2m+1)}(t_1,\ldots,t_{2m+1})=0,\qquad m=0,1,2,...
\end{equation}
This vanishing is an exact consequence of the sign-reversal symmetry, rather than a small-asymmetry approximation \cite{Schriefl2006,Galperin2006}.}
\red{
Sign-reversal symmetry alone does not imply Gaussian statistics. For a symmetric
TLF and ordered times $t_1\leq t_2\leq t_3\leq t_4$, setting $n=4$ in
Eq.~\eqref{eq:recursion-symmetric} gives
\begin{equation}
\left\langle\delta\xi_j(t_1)\delta\xi_j(t_2)\delta\xi_j(t_3)\delta\xi_j(t_4)\right\rangle_{\bar\xi_j=0}=e^{-W_j[(t_2-t_1)+(t_4-t_3)]}.
\end{equation}
Combining this result with Eq.~\eqref{C2} and Eq.~\eqref{C4}, we have 
\begin{equation}
\tilde C_{\beta_j}^{(4)}(t_1,t_2,t_3,t_4)\Big|_{\bar\xi_j=0}=-2\lambda_j^4e^{-W_j[(t_3-t_1)+(t_4-t_2)]}.
\end{equation}
The fourth-order cumulant is therefore generally nonzero. This explicitly demonstrates that a symmetric TLF can remain non-Gaussian despite the vanishing of all odd-order cumulants. 
%The corresponding fourth-order cumulant and trispectrum are numerically reconstructed below.
}

%For ordered times $t_1\leq t_2\leq t_3$, the Markov property gives
%\begin{equation}
%\left\langle\delta\xi_j(t_1)\delta\xi_j(t_2)\delta\xi_j(t_3)\right\rangle=e^{-W_j(t_3-t_2)}
%\left\langle\delta\xi_j(t_1)\delta\xi_j^2(t_2)\right\rangle.
%\end{equation}
%we obtain
%\begin{equation}
%\left\langle\delta\xi_j(t_1)\delta\xi_j(t_2)\delta\xi_j(t_3)\right\rangle=-2\bar\xi_j\left(1-\bar\xi_j^2\right)e^{-W_j(t_3-t_1)},\qquad t_1\leq t_2\leq t_3 .
%\end{equation}
%and hence
%\begin{equation}
%\tilde C^{(3)}_{\beta_j}(t_1,t_2,t_3)=-2\lambda_j^3\bar\xi_j\left(1-\bar\xi_j^2\right)e^{-W_j(t_3-t_1)},\qquad t_1\leq t_2\leq t_3 .
%\end{equation}
%Thus, the third-order cumulant is generally nonzero for an asymmetric TLF but vanishes for a symmetric TLF.

\begin{figure}[htbp]
    \centering
    \includegraphics[width=0.8\linewidth]{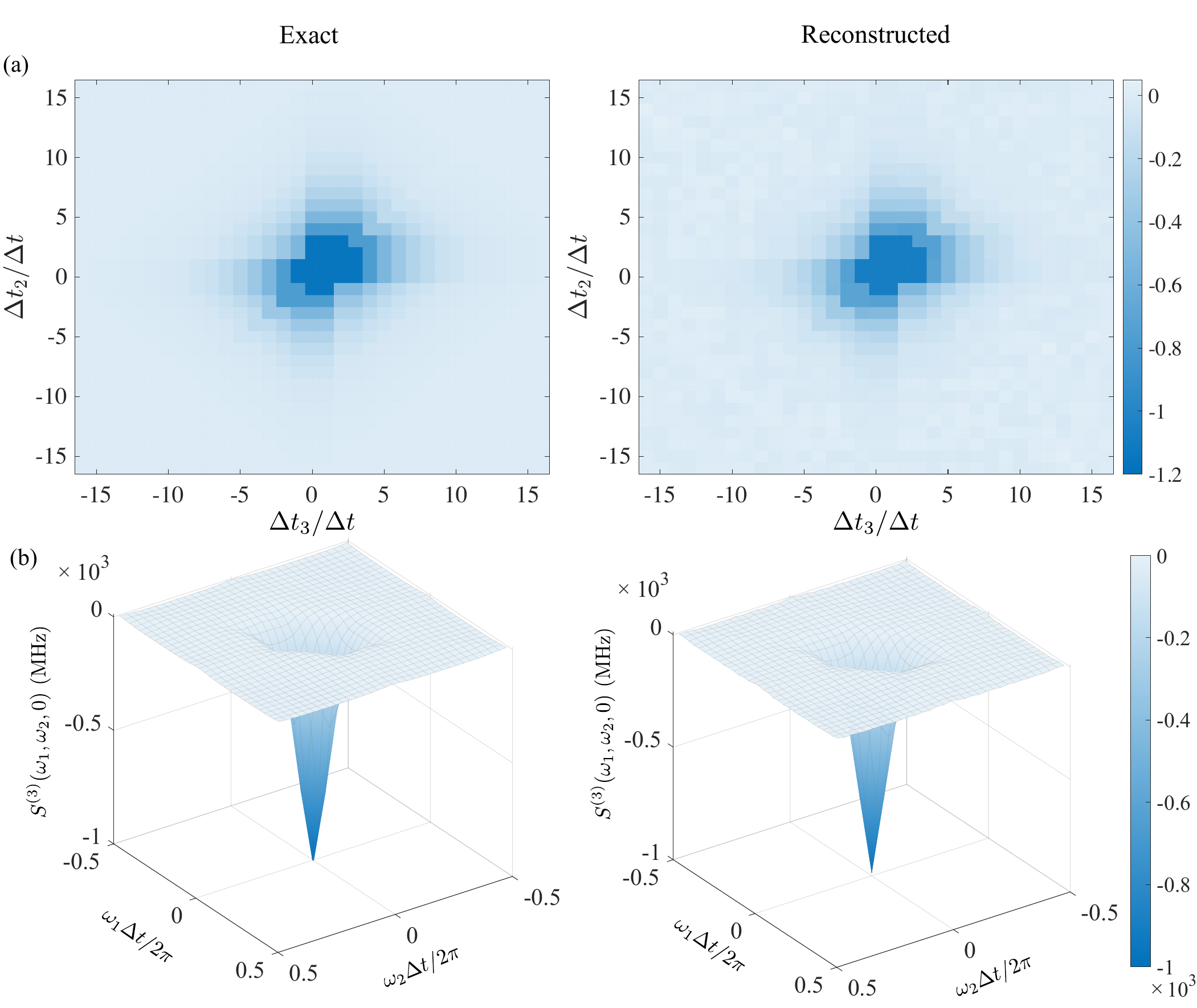}
    \caption{Reconstruction of (a) fourth-order cumulant slice $\tilde C^{(4)}(2\Delta t,\Delta t_2,\Delta t_3)$ and (b) trispectrum slice $S^{(3)}(\omega_1,\omega_2,0)$ of a classical noise. The noise is produced by a single TLF with the coupling strength $\lambda=206$ kHz and switching rate $W_j=63$ kHz. Note that the inaccessible points are smoothed by the neighborhood values for the reconstructed cumulant. The parameters are $\tau=0.25$ $\mu$s, $\Delta t$=1 $\mu$s and $N_s=5\times 10^8$.}
    \label{fig:tri}
\end{figure}

%\subsection{Reconstruction of trispectrum}
Here we show a numerical simulation of reconstructing the fourth-order cumulant and corresponding trispectrum of a single TLF \cite{ramon2019} in Fig.~\ref{fig:tri}. The reconstructed cumulant and trispectrum can recover the exact results well, demonstrating that our scheme can accurately characterize higher-order cumulants of complex classical noise processes.

\section{Effect of imperfect qubit measurement}
In this section, we show that our scheme is also feasible for non-ideal qubit measurements, such as the weak optical readout for solid-state defect spins or trapped ions \cite{degen2017a,du2024,HAFFNER2008,biercuk2009,kotler2011} and imperfect projective measurement for superconducting qubits \cite{Xiong2026}.

\subsection{Weak measurement}
We first consider the weak measurement of the qubit, which is common when the qubit is measured through an optical process. In this case, the number of detected photons in one cycle may be much smaller than one, leading to the difficulty of single-shot readout. Let $g_\alpha$ be the probability to detect one photon when the qubit is in $\ket{\alpha}$, with $\alpha=0,1$. We truncate the photon number to $n=0,1$, and the readout operators are
\begin{equation}
    \begin{aligned}
        \begin{cases}
            K_0=\sum_{\alpha=0,1}\sqrt{1-g_\alpha}P_\alpha,\\
            K_1=\sum_{\alpha=0,1}\sqrt{g_\alpha}P_\alpha.
        \end{cases}
    \end{aligned}
\end{equation}

Accordingly, the expectation value of accumulated photon number in $k$th RIM is
\begin{equation}
\ex{g_k}=\Tr(K_1U_k\ket{0}\bra{0}U_k^\dagger K_1^\dagger)=g_0p(t_k,0)+g_1p(t_k,1)
\end{equation}
with \red{$U_{k}=R_{\varphi_2}e^{-i\phi_k\sigma_z/2}R_{\varphi_1}$} and $p(t_k,\alpha_k)=[1+(-1)^{\alpha_k} \cos(\phi_k+\Delta\varphi)]/2=[1+(-1)^{\alpha_k} r_k]/2$.
Thus
\begin{equation}
    \ex{g_k}=\bar g+\Delta r_k.
\end{equation}
We define $\delta g_k=g_k-\bar g$, then we obtain the relation
\begin{equation}
    \ex{\delta g_1\cdots\delta g_n}=\Delta^n\ex{r_1\cdots r_n}\approx \Delta^n\tau^n C^{(n)}(t_1,...,t_n).
\end{equation}
Thus the weak optical readout reduces the signal by a contrast factor $\Delta^n$ but does not change the reconstructed correlation structure.

\subsection{Measurement errors}
We next consider the readout with assignment errors, where the Kraus operator becomes
\begin{equation}
    \begin{aligned}
        \begin{cases}
            M_0=\sqrt{1-p_0}P_0+\sqrt{p_1}P_1,\\
            M_1=\sqrt{p_0}P_0+\sqrt{1-p_1}P_1,
        \end{cases}
    \end{aligned}
\end{equation}
where $p_\alpha$ is the assignment error rate for state $\ket{\alpha}$. We label $m_k=+ 1(-1)$ for the outcome $\alpha_k=0(1)$, and the expectation value of $m_k$ can be calculated as
\begin{equation}
    \ex{m_k}=\Tr[(M_0^\dagger M_0-M_1^\dagger M_1)U_k\ket{0}\bra{0}U_k^\dagger]=(p_1-p_0)+(1-p_0-p_1)r_k.
\end{equation}
Let $a=p_1-p_0$, $b=1-(p_0+p_1)$ and $\delta m_k=m_k-a$, we have
\begin{equation}
    \ex{\delta m_k}=b\ex{r_k},
\end{equation}
and thus
\begin{equation}
    \ex{\delta m_1\cdots\delta m_n}=b^n\ex{r_1\cdots r_n}\approx b^n\tau^n C^{(n)}(t_1,...,t_n).
\end{equation}
So the readout errors also just affect the signal amplitude by a factor $b^n$, and do not influence our main result.

\section{Experimental acquisition time for higher-order correlation measurements}
We now estimate the required time to obtain the whole $n$-point correlation function with a superconducting qubit or a low temperature NV center. The contrast of one RIM can be written as
\begin{equation}
    G=G_{\rm ctrl}G_{\rm read}e^{-\tau/T_2}.
\end{equation}
If each $\pi/2$ pulse has an average error $r_{\pi/2}$, we take $G_{\rm ctrl}=(1-2r_{\pi/2})^2$. For approximately symmetric readout errors, $G_{\rm read}=2F_{\rm read}-1$, in which $F_{\rm read}$ is the average readout fidelity. We consider a trajectory spanning $N=48$ sampling intervals. The required sample number is calculated as in the main text, and the corresponding acquisition time is estimated by
\begin{equation}
    T_{\rm tot}\approx N N_s t_{\rm cyc},
\end{equation}
where $t_{\rm cyc}$ is the duration of one complete cycle. To simplify the calculation, below we take $A_n=0.2^n$, $\tau=0.25\,\mu$s, $\eta=0.1$, and $1-\epsilon=0.95$. 

The estimate above treats each sequence of $N$ RIMs as one independent trajectory. For a stationary noise process, the acquisition time can be reduced by sliding-window averaging. A longer trajectory containing $M>N$ sequential RIMs has effectively $S=M-N+1$ samples of length-$N$ sequence. Suppose $N_t$ length-$M$ trajectories are acquired, for example, the two-point correlation function can be estimated by 
\begin{equation}
	C^{(2)}(l_1\Delta t,l_2\Delta t)=C^{(2)}[(l_2-l_1)\Delta t]\approx\frac{1}{\tau^2 N_tS}\sum_{i=1}^{N_t}\sum_{m=1}^{S}(-1)^{\alpha^{(i)}_{m}}(-1)^{\alpha^{(i)}_{m+l_2-l_1}}.
\end{equation}
 To retain a total of $N_s$ samples, the required number of such length-$M$ trajectories is approximately $N_t\approx N_s/(M-N+1)$, giving
\begin{equation}\label{eq:ttot}
    T_{\rm tot}\approx \frac{M}{M-N+1}N_s t_{\rm cyc}.
\end{equation}
 For $M\gg N$, this time approaches $N_s t_{\rm cyc}$, corresponding to a reduction by up to a factor of $N$. 
 
\subsection{Superconducting qubits}
Representative transmon experiments have demonstrated a 22-ns $\pi/2$ gate achieving an average gate error $r_{\pi/2}=7.42\times 10^{-5}$ and $T_2=204\,\mu$s \cite{Li2023}. A recent multiplexed readout experiment used a total measurement window of 250 ns and obtained $F_{\rm read}=99.61\%$ \cite{Xiong2026}. Then the contrast can be calculated as
\begin{equation}
    G=(1-2\times7.42\times 10^{-5})^2(2\times 0.9961-1)e^{-0.25/204}\approx0.991.
\end{equation}
Using $N_s \geq 2\ln(2/\epsilon)/(\eta^2G^{2n}A_n^2)$, we obtain $N_s\approx4.8\times10^5$, $1.2\times10^7$, and $3.1\times10^8$ for $n=2$, $3$, and $4$, respectively. Taking a cycle time $t_{\rm cyc}=1\,\mu$s, we have $T_{\rm tot}\approx23$ s, $9.7$ min, and $4.1$ h, respectively. In the ideal sliding-window limit $M\gg N$, Eq.~\eqref{eq:ttot} gives the theoretical minimum acquisition times $T_{\min}\approx0.48$ s, $12$ s, and $5.2$ min for $n=2$, $3$, and $4$, respectively.

\subsection{NV centers}
At $3.7$ K, single-shot readout of an NV electron spin with an average fidelity of $95\%$ and a readout/initialization time of approximately $10\,\mu$s has been demonstrated \cite{Abobeih2018}. Because $\tau$ is much shorter than the reported second-scale coherence time, the decoherence factor is close to unity. Ref. \cite{Bartling2025} reported single-qubit gate fidelities of up to $99.991(4)\%$ , corresponding to $r_{\pi/2}\approx1.0\times10^{-4}$, thus the resulting contrast is
\begin{equation}
    G\approx(1-2\times10^{-4})^2(2\times0.95-1)\approx0.900.
\end{equation}
Taking $t_{\rm cyc}=10\,\mu$s, the same calculation gives $N_s\approx7.0\times10^5$, $2.2\times10^7$, and $6.7\times10^8$ for $n=2$, $3$, and $4$, respectively. For $N=48$, the corresponding acquisition times are approximately $5.6$ min, $2.9$ h, and $3.7$ days. In the ideal sliding-window limit $M\gg N$, the theoretical minimum acquisition times from Eq.~\eqref{eq:ttot} are $T_{\min}\approx7.0$ s, $3.6$ min, and $1.9$ h, respectively. These estimates show that two- and three-point correlations are experimentally realistic, while four-point correlations require a considerably longer acquisition time.

\bibliography{ref}